 \documentstyle[preprint,eqsecnum,prb,aps]{revtex}

\begin{document}
\widetext


\draft
\preprint{}
\title{Vortex structure in $d$-wave superconductors}
\author{M. Ichioka}
\address{Department of Physics, Kyoto University,
         Kyoto 606-01, Japan}
\author{N. Hayashi, N. Enomoto and K. Machida}
\address{Department of Physics, Okayama University,
         Okayama 700, Japan}
\date{December 8, 1995}
\maketitle
\begin{abstract}
    Vortex structure of pure $d_{x^2-y^2}$-wave superconductors
is microscopically analyzed in the framework of
the quasi-classical Eilenberger equations.
    Selfconsistent solution for the $d$-wave pair potential is
obtained for the first time in the case of an isolated vortex.
    The vortex core structure, i.e., the pair potential, the supercurrent
and the magnetic field, is found to be fourfold symmetric even in the
case that the mixing of $s$-wave component is absent.
    The detailed temperature dependences of these quantities are calculated.
    The fourfold symmetry becomes clear when temperature is decreased.
    The local density of states is calculated for the selfconsistently
obtained pair potential.
    From the results, we discuss the flow trajectory of the quasiparticles
around a vortex, which is characteristic in
the $d_{x^2-y^2}$-wave superconductors.
    The experimental relevance of our results
to high temperature superconductors is also given.
\end{abstract}

\section{Introduction}
\label{sec:1}

    A number of investigations were
carried out theoretically and experimentally to identify
the symmetry of pairing state in high-$T_c$ superconductors.
    Although precise pairing symmetry has not been determined yet,
it is recognized that $d_{x^2-y^2}$-wave symmetry is most probable.
\cite{m2s}
    Recently, the vortex structure of the $d$-wave
superconductors attracts much attention
because it may have different structure from that of
conventional $s$-wave superconductors.
    The internal degrees of freedom in $d_{x^2-y^2}$-wave pairing, i.e.,
$\hat k_x^2 - \hat k_y^2$ in reciprocal space, has fourfold symmetry.
 It is expected that reflecting the symmetry,
the core structure of isolated vortex
may break the circular symmetry and show fourfold symmetry in
$d_{x^2-y^2}$-wave superconductors.
    In conventional $s$-wave superconductors, the winding number
of the pair potential is 1 around the vortex.
    In $d_{x^2-y^2}$-wave superconductors, other components of
winding number $4n+1$ ($n$: integer) may be induced and
make the vortex structure fourfold symmetry.

    Fourfold symmetric vortex in $d_{x^2-y^2}$-wave superconductors
was so far mainly considered by Ren et al.\cite{ren,xu} and
Berlinsky et al.\cite{ber} on the two-component Ginzburg-Landau (GL)
theory for $s$- and $d$-wave superconductivity.
    According to the consideration based on the two-component GL theory,
it is possible that the $s$-wave component is coupled with the $d$-wave
component through the gradient terms.
    Therefore, the $s$-wave component may be  induced when the $d$-wave
order parameter spatially varies, such as near the vortex or interface
under certain restricted conditions.\cite{xu,matsumoto}
    The induced $s$-wave component around the vortex is fourfold symmetric.
    The resulting vortex structure in $d$-wave superconductors, therefore,
exhibits fourfold symmetry.
    The structure of the induced $s$-wave component was also
analyzed by solving the tight-binding Bogoliubov-de Gennes equation
selfconsistently on a $16 \times 16$ lattice \cite{soi}
and by the quasi-classical Eilenberger theory. \cite{ich2}

    However, this fourfold anisotropy clearly appears only
when the amplitude of the induced $s$-wave component is comparable
to the $d_{x^2-y^2}$-wave component.
    It is still not clear both theoretically and experimentally
whether or not this is indeed the case of high temperature superconductors.
    In the case where the $s$-wave pairing interaction is negligibly small
compared with the $d_{x^2-y^2}$-wave one,
the induced $s$-wave order parameter $\Delta_s$ is negligible.
    In the limit $\Delta_s \rightarrow 0$, the GL equation reduces to
an equation for only $d_{x^2-y^2}$-wave component,
\begin{equation}
-(\partial_x^2+\partial_y^2)\Delta_{x^2-y^2} -\Delta_{x^2-y^2}
+|\Delta_{x^2-y^2}|^2 \Delta_{x^2-y^2} =0
\label{eq:1.1}
\end{equation}
in the usual dimensionless form.
    Since Eq.(\ref{eq:1.1}) is same as that of the conventional
$s$-wave superconductors, the vortex structure remains circular
symmetric within the GL framework.

   Strictly speaking, the GL equation (\ref{eq:1.1}) is valid
only near the transition temperature.
   Far from the transition temperature, we have to include several
correction terms.
   Among them, there are some terms leading to  fourfold symmetry,
such as the so-called non-local correction term, \cite{ich2,taka}
\begin{equation}
-\gamma  \ln\left({T_c \over T}\right)
\biggl\{{7 \over 8}(\partial_x^2 +\partial_y^2)^2
           -\partial_x^2 \partial_y^2 \biggr\} \Delta_{x^2-y^2},
\label{eq:1.2}
\end{equation}
where $\gamma=62\zeta(5)/49\zeta(3)^2$.
    Including these terms neglected in the conventional GL theory,
we can investigate the fourfold symmetric vortex in pure $d$-wave
superconductors.
    On lowering temperature, these correction terms make important role
and fourfold symmetry is expected to be evident.

    The purpose of this paper is to investigate the vortex
structure in pure $d_{x^2-y^2}$-wave superconductors
using the quasi-classical Eilenberger equations,  \cite{eil}
and show that fourfold symmetric vortex is realized even in the
case the induced $s$-wave component is absent, i.e.,
in the case that the $s$-wave pairing interaction can be neglected.
    Merits of the quasi-classical calculation are as follows.
    It can be applied at arbitrary temperatures,
while the GL theory is valid only near the transition temperature.
    We can consider the terms neglected in the GL theory,
the higher order terms of order parameter and the higher order
derivative terms.

    The quasi-classical calculations on the vortex structure
were selfconsistently carried out for conventional $s$-wave superconductors
by Pesch and Kramer,\cite{pesch} and Klein\cite{klein} quite thoroughly.
    For pure $d_{x^2-y^2}$-wave superconductors,
the quasi-classical calculations on the vortex were so far performed
by assuming circular symmetry for the amplitude of the
$d_{x^2-y^2}$-wave pair potential.
    Schopohl and Maki showed that local density of states
exhibits characteristic fourfold symmetric distribution. \cite{sch}
    The current authors demonstrated that the supercurrent and magnetic field
distribution around an isolated vortex are fourfold symmetric.\cite{ich1}
    These results suggest that the pair potential itself is fourfold
symmetric although the circular symmetry is assumed in the initial
pair potential.
    In this paper, the selfconsistent calculation for the
$d_{x^2-y^2}$-wave pair potential is performed for the first time.
    The fourfold symmetry of vortex structure, i.e.,
the pair potential, the supercurrent and the magnetic field distribution
around a vortex is demonstrated to exist and the  temperature dependence of
these quantities is  investigated.
    We also calculate the local density of states in detail based on
the selfconsistently obtained pair potential.

     Here we consider the case of an isolated vortex
 under a magnetic field applied parallel
to the $c$-axis (or $z$-axis) in the clean limit.
    The Fermi surface is assumed to be two-dimensional,
which is appropriate to high-$T_c$ superconductors,
and isotropic for simplicity.
    Throughout the paper, energies and lengths are measured
in units of the uniform gap $\Delta_0$ at $T=0$ and
the coherence length $\xi_0=v_F/\Delta_0$ ($v_F$: Fermi velocity),
respectively.

    The rest of this paper is organized as follows.
    In Sec. \ref{sec:2}, we describe the method of calculation based
on the quasi-classical Eilenberger theory.
    Section \ref{sec:3} contains selfconsistent numerical results
of the vortex structure computed at various temperatures.
    The summary and discussions are given in Sec. \ref{sec:4}.

\section{Quasi-classical Eilenberger theory}
\label{sec:2}

To obtain the quasi-classical Green functions,
we solve the quasi-classical Eilenberger equations
for a pair potential,
\begin{equation}
\Delta(\theta,{\bf r})
= \bar\Delta(\theta,{\bf r}) e^{i\phi} ,
\label{eq:2.1}
\end{equation}
where $r=|{\bf r}|=\sqrt{x^2+y^2}$ is the distance
from the center of a vortex line situated at the origin,
$\theta$ is the angle of ${\bf k}$-vector
with the $a$-axis (or $x$-axis),
and $e^{i\phi} =(x+iy)/r$ is factored out for later
convenience.
    After an appropriate gauge transformation about $\phi$,
the Eilenberger equations for the quasi-classical Green functions
with the Matsubara frequency $\omega_n=(2n+1)\pi T$
are given as
\begin{equation}
\Bigl\{ \omega_n +{1 \over 2}
\Bigl(\partial_\parallel
+ i\partial_\parallel \phi \Bigr) \Bigr\}{\bar f}
(\omega_n,\theta,{\bf r})
= \bar\Delta(\theta,{\bf r}) g(\omega_n,\theta,{\bf r}) ,
\label{eq:2.2}
\end{equation}
\begin{equation}
\Bigl\{ \omega_n -{1 \over 2}
\Bigl(\partial_\parallel
-i\partial_\parallel \phi \Bigr) \Bigr\}
{\bar f}^\dagger(\omega_n,\theta,{\bf r})
= \bar\Delta^\ast(\theta,{\bf r}) g(\omega_n,\theta,{\bf r}),
\label{eq:2.3}
\end{equation}
\begin{equation}
\partial_\parallel g(\omega_n,\theta,{\bf r})
=\bar\Delta^\ast(\theta,{\bf r}) {\bar f}(\omega_n,\theta,{\bf r})
-\bar\Delta(\theta,{\bf r}) {\bar f}^\dagger(\omega_n,\theta,{\bf r}),
\label{eq:2.4}
\end{equation}
\begin{equation}
g(\omega_n,\theta,{\bf r})
=\Bigl(1-{\bar f}(\omega_n,\theta,{\bf r})
    {\bar f}^\dagger(\omega_n,\theta,{\bf r}) \Bigr)^{1/2}, \quad
{\rm Re} g(\omega_n,\theta,{\bf r}) > 0 ,
\label{eq:2.5}
\end{equation}
where $ \partial_\parallel = d/dr_\parallel $ and
$\partial_\parallel \phi= -r_\perp /r^2$.
    Here, we have taken the coordinate system:
$\hat{\bf u}=\cos\theta \hat{\bf x}+\sin\theta \hat{\bf y}$,
$\hat{\bf v}=-\sin\theta \hat{\bf x}+\cos\theta \hat{\bf y}$,
thus a point ${\bf r}=x \hat{\bf x}+y \hat{\bf y}$
is denoted as
${\bf r}=r_\parallel \hat{\bf u}+r_\perp \hat{\bf v}$.
    The anomalous Green functions
${\bar f}$ and ${\bar f}^\dagger$
in Eqs.(\ref{eq:2.2})-(\ref{eq:2.5}) are related to
the usual notations $f$ and $f^\dagger$ as
$f={\bar f}e^{i\phi}$ and
$f^\dagger={\bar f}^\dagger e^{-i\phi}$.
    Since we consider the isolated vortex in the extreme type II
superconductors, the vector potential in Eqs.(\ref{eq:2.2}) and
(\ref{eq:2.3}) can be neglected.

    We solve the first-derivative equations (\ref{eq:2.2})-(\ref{eq:2.4})
along the trajectory where $r_\perp$ is held constant.
    To obtain the quasi-classical Green functions,

\begin{equation}
\hat g \equiv
\left( \begin{array}{cc} g & i \bar f \\
-i \bar f^\dagger &-g \end{array}\right),
\label{eq:2.6}
\end{equation}
we use the so-called explosion method.\cite{klein,thuneberg}
    In addition to a physical solution $\hat g_{\rm ph}$,
Eqs. (\ref{eq:2.2})-(\ref{eq:2.4}) have two unphysical solutions
$\hat g_+ $ and $\hat g_-$.
    The solutions $\hat g_\pm$ explode (increase exponentially)
in the directions $\pm {\bf k}$ and decrease in the opposite directions.
    Even when we use the physical solution as a initial value,
the unphysical solutions always mix and become dominant during
the process of the numerical integration of Eqs. (\ref{eq:2.2})-(\ref{eq:2.4})
along a long path.
    We obtain $\hat g_\pm$ by integrating
from $r_\parallel \mp r_{\rm A}$ to $ r_\parallel $,
where $r_{\rm A}(>0)$ is large so that explosion takes place.
    It is known\cite{klein,thuneberg} that the physical solution
is obtained from the commutator of the two unphysical solutions,
\begin{equation}
\hat g_{\rm ph}= c [ \hat g_+ , \hat g_- ],
\label{eq:2.7}
\end{equation}
where $c$ is a constant determined from Eq.(\ref{eq:2.5}).

    The quasi-classical Green functions are also obtained
by the method of the Riccati transformation.
    The transformation removes the unphysical solutions
in the numerical integration.
    Using the parameterization devised by Schopohl and Maki,\cite{sch}
\begin{equation}
{\bar f} = {2 {\bar a} \over 1+{\bar a}{\bar b}} , \quad
{\bar f}^\dagger = {2 {\bar b} \over 1+{\bar a}{\bar b}} , \quad
g={1-{\bar a}{\bar b} \over 1+{\bar a}{\bar b} } ,
\label{eq:2.8}
\end{equation}
in Eqs. (\ref{eq:2.2})-(\ref{eq:2.4}),
we obtain the Riccati equations
\begin{equation}
\partial_\parallel {\bar a}(\omega_n,\theta,{\bf r})
=\bar\Delta(\theta,{\bf r})
-\Bigl\{2 \omega_n + i\partial_\parallel \phi
+ \bar\Delta^\ast(\theta,{\bf r}) {\bar a}(\omega_n,\theta,{\bf r})
\Bigr\}{\bar a}(\omega_n,\theta,{\bf r}),
\label{eq:2.9}
\end{equation}
\begin{equation}
\partial_\parallel {\bar b}(\omega_n,\theta,{\bf r})
=-\bar\Delta^\ast(\theta,{\bf r})
+\Bigl\{2 \omega_n + i\partial_\parallel \phi
+ \bar\Delta(\theta,{\bf r}) {\bar b}(\omega_n,\theta,{\bf r})
\Bigr\}{\bar b}(\omega_n,\theta,{\bf r}).
\label{eq:2.10}
\end{equation}
    We integrate Eqs. (\ref{eq:2.9}) and (\ref{eq:2.10})
using the solution far from the vortex
\begin{equation}
\bar a_\infty = {\sqrt{\omega_n^2 +|\bar\Delta(\theta,{\bf r})|^2 }
-\omega_n \over \bar\Delta^\ast(\theta,{\bf r})}, \quad
\bar b_\infty = {\sqrt{\omega_n^2 +|\bar\Delta(\theta,{\bf r})|^2 }
-\omega_n \over \bar\Delta(\theta,{\bf r})} ,
\quad (\omega_n > 0)
\label{eq:2.11}
\end{equation}
as an initial value, and obtain $\bar a, \bar b$
and the quasi-classical Green functions.
    We confirm that both methods, the explosion method and the Riccati
transformation method, give the same solution.

    The selfconsistent condition is given as
\begin{equation}
\bar\Delta(\theta,{\bf r})=N_0
2 \pi T \sum_{\omega_n>0} \int_0^{2\pi}{d\theta \over 2\pi}
V(\theta,\theta'){\bar f}(\omega_n,\theta',{\bf r}) ,
\label{eq:2.12}
\end{equation}
where $N_0$ is the density of states at the Fermi surface.
The pair potential and the pairing interaction are
decomposed into $s$-, $d_{x^2-y^2}$- and $d_{xy}$-wave
components,
\begin{equation}
V({\bf \theta,\theta'})
=V_s + V_{x^2-y^2} \cos(2\theta)\cos(2\theta')
+ V_{xy} \sin(2\theta)\sin(2\theta') ,
\label{eq:2.13}
\end{equation}
\begin{equation}
\bar\Delta(\theta,{\bf r})=\bar\Delta_s({\bf r})
+ \bar\Delta_{x^2-y^2}({\bf r})\cos(2\theta)
+ \bar\Delta_{xy}({\bf r})\sin(2\theta)  .
\label{eq:2.14}
\end{equation}
    Substituting Eqs.(\ref{eq:2.13}) and (\ref{eq:2.14})
into Eq.(\ref{eq:2.12}), we obtain the following selfconsistent
equations for each component:
\begin{equation}
\bar\Delta_s({\bf r})=V_s N_0
2 \pi T \sum_{\omega_n>0} \int_0^{2\pi}{d\theta \over 2\pi}
{\bar f}(\omega_n,\theta,{\bf r}) ,
\label{eq:2.15}
\end{equation}
\begin{equation}
\bar\Delta_{x^2-y^2}({\bf r})=V_{x^2-y^2} N_0
2 \pi T \sum_{\omega_n>0} \int_0^{2\pi}{d\theta \over 2\pi}
{\bar f}(\omega_n,\theta,{\bf r}) \cos(2\theta) ,
\label{eq:2.16}
\end{equation}
\begin{equation}
\bar\Delta_{xy}({\bf r})=V_{xy}N_0
2 \pi T \sum_{\omega_n>0} \int_0^{2\pi}{d\theta \over 2\pi}
{\bar f}(\omega_n,\theta,{\bf r}) \sin(2\theta) .
\label{eq:2.17}
\end{equation}
    In Eqs. (\ref{eq:2.15}) and (\ref{eq:2.17}),
$\bar\Delta_s({\bf r})$ and $\bar\Delta_{xy}({\bf r})$
are proportional to $V_s$ and $V_{xy}$, respectively.
    Here, we consider the case $|V_s|,|V_{xy}| \ll V_{x^2-y^2}$
as mentioned before.
    In this case, since $\bar\Delta_s({\bf r})$ and
$\bar\Delta_{xy}({\bf r})$ are  negligible compared with
$\bar\Delta_{x^2-y^2}({\bf r})$,
the $s$- and $d_{xy}$-wave components
do not affect $d_{x^2-y^2}$-wave superconductivity.
    For the case $V_s$ and $V_{xy}$ are negligible but finite,
the structure of $\bar\Delta_s({\bf r})$ and $\bar\Delta_{xy}({\bf r})$
are calculated from Eqs. (\ref{eq:2.15}) and (\ref{eq:2.17})
respectively, which was in detail reported in our previous paper.\cite{ich2}

    From now on, we set $\bar\Delta_s({\bf r})= \bar\Delta_{xy}({\bf r})=0$.
    Equation (\ref{eq:2.16}) is the selfconsistent condition
for $d_{x^2-y^2}$-wave superconductivity.
    In our calculation, we use the relation
\begin{equation}
{2 \over V_{x^2-y^2}N_0}=\log {T \over T_c}
+2\pi T \sum_{0<\omega_n < \omega_c} {1 \over |\omega_n|},
\label{eq:2.18}
\end{equation}
and set the energy cutoff: $\omega_c=20 T_c$.
    We calculate the r.h.s. of Eq. (\ref{eq:2.16}) and obtain the
new value for $\bar\Delta_{x^2-y^2}({\bf r})$.
    Using the renewed pair potential, we solve the Eilenberger equations
(\ref{eq:2.2})-(\ref{eq:2.5}) again.
    Starting from the initial form
\begin{equation}
\bar\Delta_{x^2-y^2}({\bf r})=\bar\Delta(T)\tanh r ,
\label{eq:2.19}
\end{equation}
we repeat this simple iteration procedure twenty times and obtain
a sufficiently selfconsistent solution for $\bar\Delta_{x^2-y^2}({\bf r})$.
    Here, $\bar\Delta(T)$, the temperature-dependent uniform gap
without magnetic field, is obtained from the BCS relation.

    In our numerical calculations, we discretize ${\bf r}$
in the region $r<r_c=10$ for $\bar\Delta_{x^2-y^2}({\bf r})$.
    In order to avoid the spurious symmetry breaking due to
computational artifacts, the mesh points are located on the cylindrical
coordinate so that the choice of mesh points is designed  not to break
cylindrical symmetry.
    When we solve Eqs. (\ref{eq:2.2})-(\ref{eq:2.4}), we need to know
$\bar\Delta_{x^2-y^2}({\bf r})$ for arbitrary {\bf r}.
    It is given by interpolation of the values on the mesh points.
    Far from the vortex $r>r_c$, we assume the pair potential to be
the form of Eq. (\ref{eq:2.19}).
    To confirm the validity of this assumption, we calculate also
for the other $r_c$ or other asymptotic form, such as
$\bar\Delta(T)\{1-(2r^2)^{-1}\}$.
    These changes do not affect the final results of the vortex
core structure.

    The supercurrent around a vortex is given in terms of
$ g(\omega_n,\theta,{\bf r}) $ by
\begin{equation}
 {\bf J}({\bf r})=2ev_F N_0 2\pi T \sum_{\omega_n>0}
\int_0^{2\pi}{d\theta \over 2\pi}{\hat{\bf k} \over i}
g(\omega_n,\theta,{\bf r}) .
\label{eq:2.20}
\end{equation}
    The associated magnetic field distribution is determined through
\begin{equation}
 {4\pi \over c}{\bf J}({\bf r})=\nabla\times {\bf H}({\bf r}) .
\label{eq:2.21}
\end{equation}
    The local density of states is given by
\begin{equation}
 N({\bf r},E)=\int_0^{2\pi}{d\theta \over 2\pi}
{\rm Re}\ g(i\omega_n \rightarrow E+i\eta,\theta,{\bf r}) ,
\label{eq:2.22}
\end{equation}
where $\eta$ is a positive infinitesimal constant.
    To obtain $ g(i\omega_n \rightarrow E+i\eta,\theta,{\bf r})$,
we solve Eqs.(\ref{eq:2.2})-(\ref{eq:2.5}) for $\eta-iE$ instead of $\omega_n$
using the selfconsistently obtained pair potential.

\section{Fourfold symmetric vortex structure}
\label{sec:3}
\subsection{Pair potential}

    We calculate the vortex structure at $T/T_c=0.1 \sim 0.7$.
    First, we consider the $d_{x^2-y^2}$-wave pair potential.
    The pair potential $\bar\Delta_{x^2-y^2}({\bf r})$ at
$T/T_c=0.1$ is shown in Fig. \ref{fig:1}.
    The amplitude of the pair potential clearly exhibits fourfold symmetry
as shown in Fig. \ref{fig:1} (a).
    The amplitude along the $0^\circ$ direction
(along the $x$-axis and $y$-axis)
is suppressed compared with that along the $45^\circ$ direction
(along the lines $y=\pm x$).
    The phase of the pair potential also exhibits fourfold symmetry,
as shown in Fig. \ref{fig:1} (b).
    In conventional $s$-wave superconductors,
since $\bar\Delta$ is real, $\arg\Delta=\arg(\bar\Delta e^{i\phi})=\phi$,
that is, the phase increases uniformly around a vortex.
 In contrast, the phase does not
increase uniformly in $d_{x^2-y^2}$-wave superconductors.
    The phase $\arg\Delta$ is larger than $\phi$ in the region
$0<\phi<\pi/4$, and smaller than $\phi$ in the region $\pi/4<\phi<\pi/2$.
    These behaviors of the pair potential can be explained as follows
near the vortex center, where the pair potential can be expanded
in the odd power of $r$.
    For the case of the fourfold symmetry, the pair potential
consists of terms with $e^{i(4n+1)\phi}$ ($n$: integer).
    Following the well-known consideration about vortex core,\cite{parks}
the small $r$-expansion begins from the order $r^{|4n+1|}$ for the factor of
$e^{i(4n+1)\phi}$.
    The pair potential is, therefore, given as follows
near the vortex center,
\begin{equation}
\Delta_{x^2-y^2}=\bar\Delta_{x^2-y^2}e^{i\phi}
= a_0 \{ (r +b_0 r^3)e^{i\phi} + c_0 r^3 e^{-3i\phi} +O(r^5) \} ,
\label{eq:3.1}
\end{equation}
where $a_0$, $b_0$ and $c_0$ are constants.
    From Eq. (\ref{eq:3.1}), the amplitude and phase of the pair potential
are, respectively, obtained as follows,
\begin{equation}
|\Delta_{x^2-y^2}| = a_0 \{ r+ b_0 r^3+ c_0 r^3 \cos 4\phi +O(r^5) \} ,
\label{eq:3.2}
\end{equation}
\begin{equation}
{\rm arg}\bar\Delta_{x^2-y^2}= -c_0 r^2 \sin 4 \phi +O(r^4)  \\ .
\label{eq:3.3}
\end{equation}
    The case $c_0<0$ is consistent with our numerical calculations
shown in Fig. \ref{fig:1}.
    Far from the vortex, the fourfold symmetry gradually fades
and reduces to circular symmetry.

    The temperature dependence of $\bar\Delta_{x^2-y^2}({\bf r})$
is presented in Fig. \ref{fig:2}.
    Figure \ref{fig:2} (a) is for the amplitude of the pair potential
as a function of $r$ along the $0^\circ$ direction,
$|\bar\Delta_{x^2-y^2}(r,\phi=0)|$.
    It shows that the size of the vortex core steeply decreases
on lowering temperature.
    Also in the conventional $s$-wave case, this reduction was reported
by Pesch and Kramer,\cite{pesch} and Gygi and Schl\"uter.\cite{gygi}
    Concerning to the temperature dependence of fourfold symmetry,
Fig. \ref{fig:2} (b) shows the difference of the amplitude along
the $0^\circ$ and the $45^\circ$ direction,
$|\bar\Delta_{x^2-y^2}(r,\phi=0)|-|\bar\Delta_{x^2-y^2}(r,\phi=\pi/4)|$.
    And Fig. \ref{fig:2} (c) shows the phase of $\bar\Delta_{x^2-y^2}$
along the $22.5^\circ$ direction, $\arg\bar\Delta_{x^2-y^2}(r,\phi=\pi/8)$.
    From both figures, we see that fourfold symmetry of vortex
becomes clear on lowering temperature.
    To examine these temperature dependences quantitatively,
the curves in Figs. \ref{fig:2} (a) and (b) are fitted by the following
functions respectively in the core region $r<r_0/3$, where $r_0$ is
defined as the radius $r$ giving minimum in Fig. \ref{fig:2} (b)
for each curves,
\begin{equation}
|\bar\Delta_{x^2-y^2}(r,\phi=0)|=\bar\Delta(T)\left\{
{r \over \xi}
+ a_3 \left({r \over \xi}\right)^3
+ a_5 \left({r \over \xi}\right)^5
+ a_7 \left({r \over \xi}\right)^7 \right\} ,
\label{eq:3.4}
\end{equation}
\begin{equation}
|\bar\Delta_{x^2-y^2}(r,\phi=0)|-|\bar\Delta_{x^2-y^2}(r,\phi=\pi/4)|
=\bar\Delta(T)\left\{
b_1{r \over \xi}
+ b_3 \left({r \over \xi}\right)^3
+ b_5 \left({r \over \xi}\right)^5
+ b_7 \left({r \over \xi}\right)^7 \right\} ,
\label{eq:3.5}
\end{equation}
where $\xi$, $a_3$, $a_5$, $a_7$, $b_1$, $b_3$, $b_5$ and $b_7$ are
fitting parameters.
    We attain the best fit with the values summarized in Table \ref{table:1}.
    The obtained value for $b_1$ is very small and regarded as 0,
which is consistent with the consideration leading to  Eq. (\ref{eq:3.2}).
    The fitting parameter $\xi$ thus obtained characterizes the coherence
length which corresponds to the actual size of the vortex core.
    Its temperature dependence is plotted in Fig. \ref{fig:3} (a),
showing  that the core size $\xi$ linearly approaches 0
on lowering temperature.
    From Table \ref{table:1}, we can see that the contribution
of the higher order terms increases with decreasing temperature.
    Figure \ref{fig:3} (b) shows the temperature dependence of the
fitting parameters $a_3$ and $b_3$.
    When temperature is decreased, both $|a_3|$ and $|b_3|$  similarly
increase, which means that the contribution of the fourfold symmetric
$r^3$ term increases.
    On the other hand, in the limit $T \rightarrow T_c$,
the parameters approach the values expected from the GL equation,
$a_3 \rightarrow -(8 \xi_{\rm GL}^2)^{-1}$, $b_3 \rightarrow 0$.

\subsection{Supercurrent and magnetic field around a vortex}

    We show a stereographic view of the current distribution in
Fig. \ref{fig:4} (a).
    It is clear that the amplitude of the current $ |{\bf J}({\bf r})|$
is fourfold symmetric.
    There are four small peaks around the core at $45^\circ$ and
its equivalent directions.
    The breaking of the cylindrical symmetry is stronger in the core
region, while toward the outer region the cylindrical symmetry is
gradually recovered.
    To consider the direction of the current flow, we decompose the
current to the radial and the rotational components.
    The rotational component has almost the same distribution as
$ |{\bf J}({\bf r})|$ shown in Fig. \ref{fig:4} (a).
    The radial component $J_r({\bf r})$ is shown in Fig. \ref{fig:4} (b),
which indicates that the current has small fourfold symmetric radial
component.
    The counter-clockwise current has negative (positive) $J_r$ and
curves inward (outward) for $0 < \phi < \pi/4$ ($\pi/4 < \phi < \pi/2$).
    The current flow, therefore, deviates from circular trajectory,
and forms fourfold symmetric trajectory, which comes near the center of
vortex at $\phi=\pi/4$.

    The associated magnetic field distribution $H({\bf r})$ is displayed
in Fig. \ref{fig:5}.
    The contours of constant magnetic field coincide with
the supercurrent streamlines.
    Reflecting the current distribution, the magnetic field is also
fourfold symmetric.
    In the core region, the field extends along the $0^\circ$ direction,
rotated by $45^\circ$ from the current distribution.
    This is because the field is strongly screened by the induced current
along the $45^\circ$ direction.

    Next, we consider the temperature dependences of these quantities.
    The current and the magnetic field distribution are shown,
respectively, in Figs. \ref{fig:6} and \ref{fig:7} at various
temperatures.
    When temperature is lowered, the peak of the current moves toward
the core and its height increases, as shown in Fig. \ref{fig:6} (a),
and magnetic field distribution shrinks toward the core,
as shown in Fig. \ref{fig:7} (a).
    These features reflect the narrowing of the core size with
decreasing temperature as mentioned before.
    As for the anisotropy, the fourfold symmetry becomes clear
in the current, as shown in Figs. \ref{fig:6} (b) and (c), and
the magnetic field, as shown in Fig. \ref{fig:7} (b), with decreasing
temperature.
    It is seen from Fig. \ref{fig:6} (b) that
$|{\bf J}(r,\phi=0)|-|{\bf J}(r,\phi=\pi/4)|$ changes sign, positive
outside of the vortex core.
    Because of this change,  the fourfold symmetric magnetic field
becomes almost circular beyond the core region.

\subsection{Local density of states}

    The local density of states $N({\bf r},E)$ is calculated
for the selfconsistently obtained pair potential.
    We confirm that the selfconsistent calculation gives essentially
the same fourfold symmetric local density of states as that predicted
by Schopohl and Maki \cite{sch} using a test potential
$\bar\Delta_{x^2-y^2}({\bf r})=\bar\Delta(T)\tanh r$.
    Figure \ref{fig:8} shows $N({\bf r},E)$ for $E$=0.2, 0.01 and 0
where we use the selfconsistent pair potential at $T/T_c=0.1$.
    From Fig. \ref{fig:8}(a) it is recognized that $N({\bf r},E)$
for $E=0.2$ has the similar fourfold symmetric distribution to that of
Schopohl and Maki in the core region.
    Because of narrowing of the core size at low temperatures
    in our selfconsistent calculation,
the characteristic length scale for the variation of $N({\bf r},E)$ is
much shorter than that of theirs.
    Figure \ref{fig:8}(b) shows $N({\bf r},E)$ for $E$=0.2
in a wider region.
    It is noted that the eight ridges of $N({\bf r},E)$ extend to
far from the vortex.
    This feature is not covered by Schopohl and Maki.\cite{sch}
    To consider the case of lower $E$ and the limit $E \rightarrow 0$,
we show $N({\bf r},E)$ for $E=0.01$ in Fig. \ref{fig:8}(c), and
for $E=0$ in Fig. \ref{fig:8}(d).

    The distribution of $N({\bf r},E)$  can be understood as follows.
    The peak lines in Fig. \ref{fig:8} consist of four flow lines of
quasiparticle with energy $E$.
    They screen the magnetic field in total.
    The flow lines are schematically presented in Fig. \ref{fig:9},
where four lines are labeled as 1$\sim$4, respectively.
    The quasiparticle on the line 1 starts toward the $45^\circ$ direction
(the node direction of the $d_{x^2-y^2}$-wave superconductivity) at far from
the
vortex, and approaches the vortex.
    It turns around at the vortex core in a parabolic orbit, and departs from
the vortex core toward the node direction of $-45^\circ$.
    In this flow trajectory, the flow direction changes from
$45^\circ$ to $-45^\circ$ gradually so that the factor $\cos 2\theta$ of the
$d$-wave pair potential does not change its sign.
    The four flow lines of quasiparticles around a vortex  are
the characteristic point of the $d_{x^2-y^2}$-wave superconductivity,
which is contrasted with the circular flow \cite{sch}
in the conventional $s$-wave superconductivity.
    From Fig. \ref{fig:8} (b)-(d), we recognize that the flow trajectories
in Fig. \ref{fig:9} approach the center of the vortex with decreasing $E$,
and reduce to the lines $y = \pm x$ in the limit $E \rightarrow 0$.
    The $N({\bf r},E)$ for $E=0$ in Fig. \ref{fig:8} (d), therefore,
has four small ridges in the $45^\circ$ and its equivalent directions,
in addition to the usual large zero bias peak at the vortex center.
    On the other hand, when $E$ becomes larger, these small ridges
in the outer regions are invisible because the quasiparticles with
larger $E$ are in the scattering state $(E>\Delta_0\cos2\theta)$,
thus they distribute uniformly.

    In Fig. \ref{fig:10} (a), (b) and (c), we show $N({\bf r},E)$
as a function of $E$ and distance from the vortex center
along the directions $\phi=\pi /8$, $\phi=0$ and $\phi=\pi /4$ respectively.
    This type of figures is already calculated by Schopohl and
Maki\cite{sch} who evaluated it non-selfconsistently.
    We also note that Hess et al. \cite{hess} and
Renner et al.\cite{renner} performed the STM experiments to directly
observe the local density of states $N({\bf r},E)$ in a $s$-wave
superconductor 2H-NbSe$_2$ whose characteristic features are analyzed
theoretically by  Gygi and Schl\"uter,\cite{gygi} and Shore et al.\cite{shore}
    Three of the four flow trajectories in Fig. \ref{fig:9} can be seen
as three peaks for $E<1$ in Fig. \ref{fig:10} (a).
    To specify their relation, we label the three peak lines and the
corresponding positions as A, B and C in Figs. \ref{fig:9} and \ref{fig:10}
respectively.
    For $N({\bf r},E)$ along the direction $\phi=0$,
the peak lines A and B overlap each other as seen from Fig. \ref{fig:10} (b).
    With increasing $\phi$ from 0 to $\pi /4$,
peak lines A and C shift to lower energy and line B shifts to
higher energy.
    It is seen from Fig. \ref{fig:10} (c) that for $N({\bf r},E)$
along the direction $\phi=\pi /4$,
the peak lines B and C overlap each other and the line A
reduces to the peak at $E=0$.
    The peak line at $E=1$ in Fig. \ref{fig:10} corresponds to the gap edge,
which becomes clear far from the vortex.

\section{Summary and discussions}
\label{sec:4}

    The vortex structure in a pure $d_{x^2-y^2}$-wave superconductor is
studied in the framework of the quasi-classical Eilenberger equation.
    By a selfconsistent calculation, we obtain the pair potential
around an isolated vortex, which shows the characteristic fourfold
symmetry in the core region.
    The associated supercurrent and magnetic field also exhibit fourfold
symmetry around a vortex.
    We confirm that the fourfold symmetry becomes clear with decreasing
temperature.
    Using the selfconsistently obtained pair potential, the fourfold symmetric
local density of states is calculated.

    While our numerically obtained vortex structure has similar fourfold
symmetry to that predicted by Xu et al.\cite{xu} who use the two-component
GL equations, the origin of the fourfold symmetry is quite different
from theirs.
    In their theory, the fourfold symmetry is induced by the mixing of
$s$-wave component.
    On the other hand, we consider the case where the mixing of the $s$-wave
component is negligible, i.e., the pure $d_{x^2-y^2}$-wave case.
    Even in this case, the vortex structure exhibits fourfold symmetry.
    It is due to the fourfold symmetric terms which are higher order
of $\ln(T/T_c)$ and neglected  in the conventional GL theory.
    When temperature decreases below $T_c$, the contribution of these terms
increases and fourfold symmetry becomes clear.
    Since our quasi-classical calculation automatically takes
these higher order terms into account,
we can obtain the fourfold symmetry of the vortex structure even in the
pure $d_{x^2-y^2}$-wave superconductors.

    The scenario based on the two-component GL theory mentioned before
is not realized in the case the induced $s$-wave component is negligible.
    In our formulation, it corresponds to the case where the $s$-
(or extended $s$-) wave pairing interaction $V_s$ is negligible
compared with $V_{x^2-y^2}$ in Eq. (\ref{eq:2.13}).
    The induced $s$-wave component $\bar\Delta_s({\bf r})$ is
proportional to $V_s$ in Eq. (\ref{eq:2.15}).
    In such a case $|V_s| \ll V_{x^2-y^2}$, $\bar\Delta_s({\bf r})$ is
negligible and does not effect the $d_{x^2-y^2}$-wave superconductivity.
    In this case, our scenario based on the quasi-classical theory
is realized.
    The consideration of the limit $V_s \rightarrow 0$ on the GL theory
gives the same result, if the Pade approximation \cite{ren,xu} is not
used in the derivation of the GL equations.
    The coefficient of $\Delta_s$ in the GL equations is given by
$\alpha_s = (V_sN_0)^{-1}-2(V_{x^2-y^2}N_0)^{-1}$ without the Pade
approximation.
    Since $|\Delta_s| \propto \alpha_s^{-1}$, \cite{ren,xu,ber}
$\Delta_s \rightarrow 0$ in the limit $V_s \rightarrow 0$,
which is consistent with our consideration.
    On the other hand, in the GL theory with the Pade approximation,
$\alpha_s = 2(V_{x^2-y^2}N_0)^{-1}\{1+2(-V_s)/V_{x^2-y^2}\}$,
which gives an invalid result $\alpha_s \rightarrow $ finite
in the limit $V_s \rightarrow 0$.
    The Pade approximation is inadequate in the case $V_s \rightarrow 0$
and $V_s$ is repulsive.
    When $V_s$ is repulsive, the original GL theory without the
Pade approximation gives unphysical solution for $\Delta_s$.\cite{ren,xu}
    To understand the repulsive case, further careful studies are needed.
    A selfconsistent quasi-classical calculation may be one of the
possible approaches.

    The four flow trajectories of the quasiparticle for $E(<1)$, as shown in
Fig. \ref{fig:9} schematically, is the key point to understand the
fourfold symmetry of the vortex structure in a $d_{x^2-y^2}$-wave
superconductor.
    It is contrasted with the circular flow in conventional
$s$-wave superconductors.
    The local density of states in Figs. \ref{fig:8} and \ref{fig:10}
directly reflects the four flow trajectories.
    As the supercurrent is the sum of the quasiparticle flow
contributions with various $E$'s, the resulting supercurrent and
the associated magnetic field distribution around a vortex
have fourfold symmetry.
    Fourfold symmetry of the pair potential also reflects
the quasiparticle distribution.

    Reflecting the fourfold symmetry of the vortex core,
the vortex lattice or flux line lattice may be distorted
from hexagonal symmetry in $d$-wave superconductors.
    The distortion may occur at higher field as the inter-vortex distance
is short and the effect of the fourfold symmetric core structure increases.
    In the case of anisotropic core structure, we have to consider
the orientation of the vortex lattice, i.e., nearest neighbor direction of
vortices relative to the underlying crystal lattice.
     Won and Maki\cite{won} considered that at near $H_{c2}$
a square lattice tilted by $45^\circ$ from the $a$-axis is preferable
to a triangular lattice at $T \le 0.8 T_c$ in $d$-wave superconductors.
    By using the two-component GL theory mentioned above,
Berlinsky et al.\cite{ber} suggested that the Abrikosov lattice
at near $H_{c2}$ varies continuously from a triangular, through an oblique,
to a square one with increasing field and $s$-$d$ mixing parameter.
    In our consideration to the orientation,
to smoothly connect the flow trajectories of quasiparticles
in Fig. \ref{fig:9} with the nearest neighbor vortices,
the nearest neighbor direction prefers $45^\circ$,
which is consistent with the result by Won and Maki.

    In the experiment, Keimer et al.\cite{keimer} reported an oblique
lattice by a small-angle neutron scattering study of the vortex lattice
on ${\rm YBa_2Cu_3O_7}$ in a magnetic field region of 0.5T$\le H \le$5T.
    The lattice is with an angle of $73^\circ$ between the two primitive
vectors and oriented such that the nearest-neighbor direction of vortices
makes an angle of $45^\circ$ from the $a$-axis.
    The oblique lattice was also observed by scanning tunneling microscopy
(STM) by Maggio-Aprile et al.\cite{aprile}
    To understand the origin of the oblique lattice and its orientation,
in addition to the effect of the intrinsic in-plane anisotropy, that is,
the difference of the coherence lengths between $a$-axis and
$b$-axis directions,\cite{walker} the study of the vortex lattice
in $d$-wave superconductors is certainly important.

    Most probable candidate of experimental probes for detecting the
fourfold symmetry of the vortex structure is to observe the local density
of states (LDOS) by STM.
    If the four ridges placed four-fold symmetrically in LDOS
are observed at the core region, it will be possible to determine
definitely the symmetry of high-$T_c$ superconductors and distinguish
other symmetries, e.g., the four ridges of the LDOS in $d_{xy}$-wave
pairing is rotated by $45^\circ$ from that of the $d_{x^2-y^2}$-wave case,
a circular ridge of the LDOS is expected around a vortex in $s$-wave pairing.
    For the energy range $0.05 < E <1$, LDOS extends along the $x$-axis
and $y$-axis (or $a$-axis and $b$-axis) as shown in Fig. \ref{fig:8} (a).
    For $E \rightarrow 0$, LDOS extends along the lines $y= \pm x$
as shown in Fig. \ref{fig:8} (d), rotated by $45^\circ$ from that
of higher energy.
    The peak lines are detected as the line
regions of the zero-bias conductance enhancement.

    While the peak lines for $E \rightarrow 0$ was previously proposed by
Kashiwaya et al.\cite{kashiwaya} on the analogy of bound states in the
pseudo-quantum wells and in the surface region of $d_{x^2-y^2}$-wave
superconductors, our explanation of the peak lines is different from theirs.
    Their theory starts from the assumption that quasiparticles rotates
around the core along coaxial circles as in the case of conventional
$s$-wave superconductors.
    Along the circular trajectory, all quasiparticles feel the node of
the pair potential at the point $\phi=\pi/4$ and
its equivalent points, where the sign of the factor $\cos 2 \theta$ in
$d$-wave pair potential changes.
    Thus the formation of the mid-gap states are expected at these nodes,
and the bound states form two orthogonal lines $y= \pm x$.
    Following their theory, the pair potential is suppressed along
the lines $y= \pm x$.
    However, it contradicts with our numerically obtained
pair potential which is suppressed along the
$x$-axis and $y$-axis as shown in Fig. \ref{fig:1}.
    Our understanding is as follows.
    Quasiparticles flow around a vortex along not coaxial circles
but four trajectories shown in Fig. \ref{fig:9}.
    Along each of the four trajectories, quasiparticles do not feel
the sign change of the factor $\cos 2 \theta$.
    The peak lines $y= \pm x$ for $E=0$ are due to the quasiparticles
approaching and leaving the core region with the flow direction $\theta=\pi/4$
(and its equivalent directions), which is the direction energy gap is
suppressed in $d_{x^2-y^2}$-wave superconductors.
    We obtain the same LDOS also for the case when we use
$|\cos 2 \theta |$
instead of $\cos 2 \theta$, corresponding to the case
where the sign of the pair potential does not change.

    The de Haas-van Alphen (dHvA) oscillation in the mixed states is one of
the related topics with the bound states of quasiparticles around a vortex.
    The dHvA oscillation was reported also on ${\rm YBa_2Cu_3O_7}$.
\cite{smith,kido}
    For the dHvA oscillation, enough quasiparticle transfer of
low energy bound states around a vortex core should occur between vortices.
    In $d_{x^2-y^2}$-wave superconductors,
the quasiparticles with $E=0$ exists along the lines $y= \pm x$.
    Following the theory of Won and Maki,\cite{won}
the nearest neighbor vortex is located on the lines $y= \pm x$.
    Thus  the low energy bound states around each  vortex are connected
each other by the quasiparticle flow with the flow direction
$\phi=\pi/4$ and its equivalent directions.
    In this situation, dHvA oscillation is easier to occur compared with
the conventional $s$-wave case.
    This consideration can be applied to the cases of other symmetry
if the energy gap has node.

    Another topic is whether the energy level of quasiparticle bound states
is quantized around a vortex in $d$-wave superconductors.
    In conventional $s$-wave superconductors, the energy level is discretized
and the separation of the energy levels is the order of $\Delta_0^2/E_F$
($E_F$: Fermi energy).\cite{caroli}
    Wang and MacDonald\cite{wang} reported that the separation of
energy levels is visible in the $s$-wave case and is not visible
in the $d$-wave case using a short coherence length model.
    While our quasi-classical calculation can not directly consider the
quantization, our results of LDOS give the following consideration
to this problem.
    Since quasiparticles do not flow along closed circular trajectories
in $d$-wave superconductors, the quantizations could be different from
the $s$-wave case.
    For the case of open trajectories as shown in Fig. \ref{fig:9},
it is expected that the quantization is absent or, if it exists,
the separation of the energy levels is small, which seems to be consistent
with the results by Wang and MacDonald.
    To consider the quantization in detail, we have to consider
the coherence effect of quasiparticle wave functions at the cross
points of four trajectories in Fig. \ref{fig:9}.

\acknowledgments

    We would like to thank Professor Y. Tanaka
for valuable discussions.
    The authors are indebted to Supercomputer Center of
the Institute for Solid State Physics, Univ. of Tokyo
for a part of numerical calculations.


\begin{table}
\caption{
    Fitting parameters in Eqs. (3.4) and (3.5) at various temperatures
$T/T_c=0.1\sim 0.7$.
    Temperature dependent uniform gap $\bar\Delta(T)$ is also presented.
}
\begin{tabular}{ccccccccc}
$T/T_c$ & $\xi$ & $a_3$ & $a_5$ & $a_7$ & $b_3$ & $b_5$ & $b_7$ &
$\bar\Delta(T) $ \\
\hline
0.1 & 0.132 & $-$3.24  & 15.8   & $-$37.5  & $-$1.19   & 8.74  &
$-$23.7   & 1.00 \\
0.2 & 0.267 & $-$1.70  &  5.77  & $-$11.7  & $-$0.621  & 3.41  &
$-$8.00  & 0.999 \\
0.3 & 0.411 & $-$1.08  &  2.53  &  $-$4.13 & $-$0.352  & 1.46  &
$-$2.82  & 0.990 \\
0.4 & 0.558 & $-$0.782 &  1.38  &  $-$2.11 & $-$0.212  & 0.751 &
$-$1.38  & 0.971 \\
0.5 & 0.703 & $-$0.619 &  0.880 &  $-$1.38 & $-$0.131  & 0.442 &
$-$0.862 & 0.937 \\
0.6 & 0.856 & $-$0.522 &  0.648 &  $-$1.26 & $-$0.0795 & 0.285 &
$-$0.717 & 0.882 \\
0.7 & 1.03  & $-$0.468 &  0.475 &  $-$1.23 & $-$0.0492 & 0.215 &
$-$0.637 & 0.801 \\
\end{tabular}
\label{table:1}
\end{table}

\begin{figure}
\caption{
    Pair potential at $T/T_c= 0.1$.
    It is fourfold symmetric even in the case the mixing of the
$s$-wave component is absent.
    (a) Contour plot of the amplitude,
$|\bar\Delta_{x^2-y^2}({\bf r})|$.
    From the center, 0.1, 0.2, $\cdots$, 0.9.
    Amplitude is suppressed along the $0^\circ$ direction.
    (b) Stereographic view of the phase,
$\arg\bar\Delta_{x^2-y^2}({\bf r})$.
    It is positive (negative) for $0<\phi < \pi /4 $
($\pi /4 <\phi < \pi /2$).
}
\label{fig:1}
\end{figure}

\begin{figure}
\caption{
    Pair potential as a function of distance from the vortex center, $r$,
at various temperatures $T/T_c=0.1 \sim 0.7$.
    With decreasing temperature, the size of the vortex core decreases
and fourfold symmetry becomes clear.
    (a) Amplitude along the $0^\circ$ direction,
$|\bar\Delta_{x^2-y^2}(r,\phi=0)|$.
    (b) Difference of the amplitude along the $0^\circ$ and the $45^\circ$
directions, $|\bar\Delta_{x^2-y^2}(r,\phi=0)|
-|\bar\Delta_{x^2-y^2}(r,\phi=\pi/4)|$.
    (c) Phase of $\bar\Delta_{x^2-y^2}({\bf r})$ along the $22.5^\circ$
direction, $\arg\bar\Delta_{x^2-y^2}(r,\phi=\pi/8)$.
}
\label{fig:2}
\end{figure}

\begin{figure}
\caption{
    Temperature dependence of fitting parameters $\xi$ (a) and
$a_3$, $b_3$ (b).
    The lines are guides for the eye.
    Core size $\xi$ approaches 0 linearly on lowering temperature.
}
\label{fig:3}
\end{figure}

\begin{figure}
\caption{
    Current distribution around a vortex at $T/T_c= 0.1$ in units of
$-2eN_0 v_F \Delta_0$.
    (a) Stereographic view of the amplitude, $|{\bf J}({\bf r})|$.
    There are four small peaks around the core at $45^\circ$ and
its equivalent directions.
    (b) Stereographic view of the radial component, $J_r({\bf r})$.
    It is negative (positive) for $0<\phi < \pi /4 $
($\pi /4 <\phi < \pi /2$).
}
\label{fig:4}
\end{figure}

\begin{figure}
\caption{
    Contour plot of the magnetic field distribution, $H({\bf r})$,
at $T/T_c= 0.1$.
    The field extends along the $x$-axis and $y$-axis.
}
\label{fig:5}
\end{figure}

\begin{figure}
\caption{
    Current distribution as a function of $r$ at various temperatures
$T/T_c=0.1 \sim 0.7$ in units of $-2eN_0 v_F \Delta_0$.
    (a) Amplitude along the $0^\circ$ direction, $|{\bf J}(r,\phi=0)|$.
    (b) Difference of the amplitude along the $0^\circ$ and the $45^\circ$
directions, $|{\bf J}(r,\phi=0)|-|{\bf J}(r,\phi=\pi/4)|$.
    (c) Radial component along the $22.5^\circ$ direction,
$J_r(r,\phi=\pi/8)$.
}
\label{fig:6}
\end{figure}

\begin{figure}
\caption{
    Magnetic field distribution (arbitrary unit) as a function of $r$
at various temperatures $T/T_c=0.1 \sim 0.7$.
   (a) Deviation from the value of the vortex center $H_0$ along the
$0^\circ$ direction, $H(r,\phi=0)-H_0$.
   (b) Difference of the field along the $0^\circ$ and the $45^\circ$
directions, $H(r,\phi=0)-H(r,\phi=\pi/4)$.
}
\label{fig:7}
\end{figure}

\begin{figure}
\caption{
    Local density of states, $N({\bf r},E)$.
    Oscillating behavior of peak's height along the peak lines is
due to the lack of mesh points in these figures.
    If the number of mesh points is enough large, the height smoothly
varies along the peak lines.
    (a) $E=0.2$.
    Self-consistent calculation gives the similar fourfold symmetric
distribution to that of Schopohl and Maki.
    (b) The same as (a) but wider region is presented.
    It is noted that the eight peak lines extend outward from the vortex.
    (c) $E=0.01$.
    Peak lines approach the center of vortex, but have the similar
structure to (b).
    (d) $E=0$.
    In addition to the usual large zero bias peak at the vortex center,
whose height is truncated in this figure,
there are four small peak lines on the lines $y= \pm x$.
}
\label{fig:8}
\end{figure}

\begin{figure}
\caption{
    Flow trajectories of quasiparticles with energy $E(<1)$ around a vortex.
    Points A, B and C correspond to the peaks in Fig. 10.
}
\label{fig:9}
\end{figure}

\begin{figure}
\caption{
    Local density of states as a function of $E$ and $r$ along the
direction $\phi= \pi /8$ (a), $\phi= 0$ (b) and $\phi= \pi /4$ (c) .
    Three peak lines A, B and C for $E<1$ correspond to the trajectory
in  Fig. 9.
    The peak line at $E=1$ is the gap edge.
    Oscillating behavior of peak's height along the peak lines is
due to the lack of mesh points in these figures.
    If the number of mesh points is enough large, the height smoothly
varies along the peak lines.
}
\label{fig:10}
\end{figure}

\end{document}